\documentclass[a4paper,twocolumn,superscriptaddress,11pt]{quantumarticle}
\pdfoutput=1
\usepackage[utf8]{inputenc}
\usepackage[english]{babel}
\usepackage[T1]{fontenc}
\usepackage{amsmath}
\usepackage{hyperref}

\usepackage{tikz}
\usepackage{lipsum}

\newtheorem{theorem}{Theorem}

\begin{document}

\title{Function Approximation with Quantum Circuit}
\date{\today}
\author{Alberto Delgado}
\affiliation{Department of Electrical and Electronics Engineering, Universidad Nacional de Colombia, Bogotá.}
\email{adelgado@ieee.org}

\maketitle

\begin{abstract}
  A mathematical proposition with a trainable pair, operator and quantum circuit, are introduced to approximate functions expressed as cubic Taylor polynomials, numerical simulations illustrate three cases.\\
\end{abstract}

The motivations behind this paper are twofold, to explore applications of small quantum circuits \cite{preskill} and to show that sigmoid functions can be approximated with quantum circuits, it is important to mention that sigmoid functions are the building blocks of neural networks \cite{norgaard}. 
Physical quantum systems include parameterized quantum gates \cite{Q}, these gates offer the opportunity to change a quantum state with external data or even train quantum circuits to estimate unknown probability distributions \cite{benedetti}. 

The paper is organized as follows. Section one, introduces a parameterized quantum circuit and formulates a mathematical proposition to prove its function approximating capabilities. Section two, presents the simulation results to approximate three functions, quadratic, Gaussian, and sigmoid. Finally, section three is the summary.

\section{Expectation Value as Polynomial}
A mathematical proposition proves that the expectation value of an operator G, for a two qubits parameterized quantum circuit, approximates a cubic polynomial with tunable coefficients.

\subsection{Quantum Circuit}
\begin{figure}[t]
  \centering
  \includegraphics[scale=1.1]{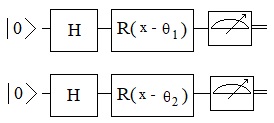}
  \caption{Function approximation with quantum circuit, $\Hat{f}(x)\hspace{1 mm} =\hspace{1mm}<G>$. Parameters $(\theta_1, \theta_2)$ and operator G are adjusted so $\lvert f(x) - \Hat{f}(x) \lvert \hspace{1 mm} \leq \hspace{1 mm}\epsilon $, $x \in [-x_0, +x_0]$.}
  \label{fig:fig1}
\end{figure}

Consider the two qubits parameterized quantum circuit, Fig.~\ref{fig:fig1}, where x is an independent variable and $(\theta_1, \theta_2)$ are parameters.
The expectation value $\Hat{f}(x)\hspace{1 mm} =\hspace{1mm}<G>$ can approximate a function $f(x)$, represented with a cubic polynomial in Taylor series, by tuning parameters $(\theta, G)$ \eqref{1} to minimize a performance index J \eqref{2} over a set of N samples in the domain $x \in [-x_0, +x_0]$.

\begin{align}
\label{1}
\theta = \begin{bmatrix} \theta_1 \\ \theta_2 \end{bmatrix}; \hspace{2 mm}
G = 
\begin{bmatrix}g_0 & 0 & 0 & 0\\
0 & g_1 & 0 & 0\\
0 & 0 & g_2 & 0\\
0 & 0 & 0 & g_3\end{bmatrix}\\
\label{2}
J = \sum \limits_{k=1}^N (f(x_k) - \Hat{f}(x_k))^2
\end{align}

\subsection{Cubic Polynomial}
It is shown that the expectation value is equivalent to a cubic polynomial by using Taylor series, the coefficients depend on parameters \eqref{1}.
\begin{theorem}
Consider the quantum state $\lvert \psi(\theta_1, \theta_2, x)>$ and operator G,
\begin{align}
\lvert \psi(\theta_1, \theta_2, x)>\hspace{1 mm} = \hspace{1 mm}U \lvert 00>\\
U = R(x - \theta_2)\otimes R(x - \theta_1) (H\otimes H)\\
R(x - \theta_i) =
\begin{bmatrix}\cos (\frac{x - \theta_i}{2}) & -\sin (\frac{x - \theta_i}{2})\\ \\ \sin (\frac{x - \theta_i}{2}) & \cos (\frac{x - \theta_i}{2}) \end{bmatrix}\\
G = 
\begin{bmatrix}g_0 & 0 & 0 & 0\\
0 & g_1 & 0 & 0\\
0 & 0 & g_2 & 0\\
0 & 0 & 0 & g_3\end{bmatrix}
\end{align}
The expectation value has the structure of a cubic polynomial,
\begin{align}
\label{5}
\Hat{f}(x)\hspace{1 mm} = \hspace{1 mm}<\psi(\theta_1, \theta_2,x) \lvert G \lvert \psi(\theta_1, \theta_2,x)>\\
\label{6}
\Hat{f}(x) = a_0+a_1x+a_2x^2+a_3x^3
\end{align}
the coefficients $a_i$ are function of parameters $ (\theta_1, \theta_2, g_0, g_1, g_2, g_3) $.\\ \\
\textit{Proof:}
\begin{enumerate}
	\item Calculate $U_1 = R(x - \theta_2) \otimes R(x - \theta_1)$
	\item Multiply $U_1$ by $(H\otimes H) \lvert 00>$
	\item Replace $\cos (\frac{x - \theta_i}{2})$ and $\sin (\frac{x - \theta_i}{2})$ with the first terms of their Taylor series.
\end{enumerate}
The amplitudes of the quantum state are quadratic polynomials, calculating the expectation value \eqref{5} results in \eqref{6}.
\end{theorem}

\section{Simulations}
A bioinspired training algorithm known as chemotaxis \cite{bremermann, delgado} and implemented in octave \cite{octave} was used to minimize the performance index \eqref{2} by tuning the set \eqref{1}. The number of samples is N = 30 in the interval with $x_0 = 1.5$.

\subsection{Quadratic function}
\begin{figure}[t]
  \centering
  \includegraphics[scale=0.75]{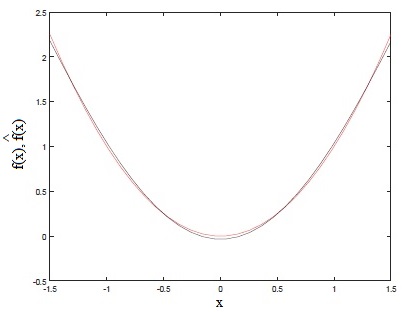}
  \caption{Quadratic function $f(x) = x^2$ (red), 5000 iterations, final J = 0.03.}
  \label{fig:fig2}
\end{figure}

Fig.~\ref{fig:fig2} shows a quadratic function $f(x) = x^2$ (red) and its approximation $\Hat{f}(x)$ (black). The final performance index was J = 0.03, parameters \eqref{9} and \eqref{10}.

\begin{align}
\label{9}
\theta &= 
\begin{bmatrix}1.373 \\ 1.770\end{bmatrix}\\
\label{10}
G &= 
\begin{bmatrix}-0.081 & 0  & 0  & 0\\
0 & 2.260 & 0 & 0\\
0 & 0 & 2.272 & 0\\
0 & 0 & 0 & 4.954\end{bmatrix}
\end{align}

\subsection{Gaussian function}
\begin{figure}[h]
  \centering
  \includegraphics[scale=0.75]{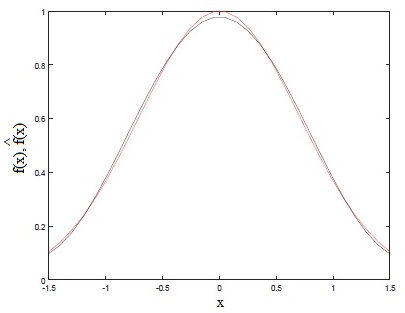}
  \caption{Gaussian function $f(x) = e^{-x^2}$ (red), after 5000 iterations, the final J = 0.005.}
  \label{fig:fig3}
\end{figure}

The second function is Gaussian $f(x) = e^{-x^2}$ shown in Fig.~\ref{fig:fig3} (red) and its approximation $\Hat{f}(x)$ (black). After training, the final set of parameters is given in \eqref{11} and \eqref{12} with a performance index \eqref{2} J = 0.005. Notice the small differences between the actual function and the approximation, there is a compromise between quantum circuit complexity and the final performance index.

\begin{align}
\label{11}
\theta &= 
\begin{bmatrix}0.497 \\ -0.498\end{bmatrix}\\
\label{12}
G &= 
\begin{bmatrix}-0.088 & 0  & 0  & 0\\
0 & 1.152 & 0 & 0\\
0 & 0 & 1.711 & 0\\
0 & 0 & 0 & -0.089\end{bmatrix}
\end{align}

\subsection{Sigmoid function}
\begin{figure}[t]
  \centering
  \includegraphics[scale=0.75]{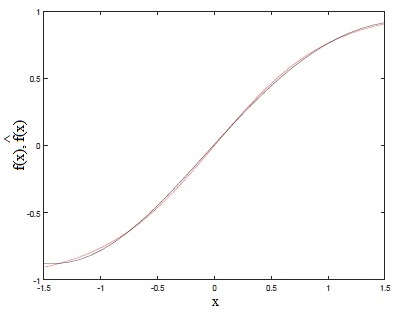}
  \caption{Sigmoid function $f(x) = tanh(x)$ (red), final J = 0.006, 5000 iterations.}
  \label{fig:fig4}
\end{figure}

Fig.~\ref{fig:fig4} shows the sigmoid function $f(x) = tanh(x)$ (red) and its approximation $\Hat{f}(x)$ (black).

The final performance index was J = 0.006, parameters \eqref{13} and \eqref{14}.

\begin{align}
\label{13}
\theta &= 
\begin{bmatrix}0.266 \\ 0.069\end{bmatrix}\\
\label{14}
G &= 
\begin{bmatrix}-0.885 & 0  & 0  & 0\\
0 & 0.055 & 0 & 0\\
0 & 0 & 0.466 & 0\\
0 & 0 & 0 & 0.931\end{bmatrix}
\end{align}

\section{Summary}
A mathematical proposition has been proved to show that the expectation value of an operator for a two qubits parameterized quantum circuit can approximate a function expressed as a third degree Taylor polynomial. Simulations for three functions, quadratic, Gaussian, and sigmoid, agree with the theory. The approximation error is related to the complexity of the quantum circuit.

Neural networks are trainable universal maps, here an operator with a parameterized quantum circuit are used to approximate a sigmoid function, the building block of neural networks. 

\bibliographystyle{plain}

\end{document}